\documentclass[12pt]{article}
\usepackage{fullpage,epsfig,graphics,amsbsy,amssymb,cancel,slashed,mathrsfs}
\usepackage{psfrag,hyperref}
\usepackage{graphicx,color}
\usepackage{wrapfig}
\usepackage{subfigure}

\newcommand{\be}{\begin{eqnarray}}
\newcommand{\ee}{\end{eqnarray}}
\usepackage{amsmath}
\usepackage{slashed}
\numberwithin{equation}{section}
\usepackage{caption}
\usepackage[nosort]{cite}

\graphicspath{{./figures/}}

\newcommand{\bP}{\mathbf{P}}
\newcommand{\bQ}{\mathbf{Q}}

\def\R{{\cal R}}

\newcommand{\bea}{\begin{eqnarray}}
\newcommand{\eea}{\end{eqnarray}}  
\newcommand{\nn}{\nonumber}
\newcommand{\Tr}{\textrm{Tr}}
\newcommand{\tr}{\textrm{tr}}

\newcommand{\NN}{\mathcal{N}}

\newcommand{\OO}{{\mathcal O}}

 \newcommand{\II}{{\mathcal I}}

\def\al{\alpha}
\def\om{\omega}

\begin{document}

\thispagestyle{empty}
\begin{flushright} \small
UUITP-15/23\\
 MIT-CTP/5573
 \end{flushright}
\smallskip
\begin{center} \LARGE
{\bf The asymptotic form of the Hagedorn temperature in planar $\NN=4$ super Yang-Mills}
 \\[12mm] \normalsize
{\bf  Simon Ekhammar${}^{a,b}$,  Joseph A. Minahan${}^{a,c}$, and Charles Thull${}^{a}$ } \\[8mm]
 {\small\it
 ${}^a$Department of Physics and Astronomy,
     Uppsala University,\\
     Box 516,
     SE-751\,20 Uppsala,
     Sweden
     
     \smallskip
     \centerline{\it and}
     
     \smallskip
    ${}^b$ Mathematics Department, King’s College London,\\
    The Strand, London WC2R 2LS, UK
     
         \smallskip
     \smallskip
     \centerline{\it and}
     
     \smallskip
     ${}^c$Center for Theoretical Physics,
     Massachusetts Institute of Technology\\
     Cambridge, MA 02139, USA
     
  }

  \medskip 
   \texttt{ \href{mailto:simon.ekhammar@physics.uu.se,joseph.minahan@physics.uu.se,charles.thull@physics.uu.se}{\{simon.ekhammar, joseph.minahan, charles.thull\}@physics.uu.se}}

\end{center}
\vspace{7mm}
\begin{abstract}
Using the supergravity dual and the  plane-wave limit as a guide, we conjecture the asymptotic large coupling form of the Hagedorn temperature for planar $\NN=4$ super Yang-Mills to order $1/\sqrt{\lambda}$.  This is two orders beyond the presently known behavior.  Using the quantum spectral curve procedure of Harmark and Wilhelm, we show that our conjectured form is in excellent agreement with the numerical results.

\end{abstract}

\eject
\normalsize

\tableofcontents

\section{Introduction}

Confining gauge theories at low temperatures can behave like string theories.  A classic sign of this is a density of states that grows as $\rho(E)\sim e^{E/T_H}$, where $T_H$ is the Hagedorn temperature.  At $T=T_H$ the string theory picture should break down.  This is often associated with the deconfinement of the gauge theory, but depending on the circumstances the deconfinement temperature is often below $T_H$ \cite{Aharony:2003sx,Aharony:2005bq}.  

By putting the gauge theory on a compact spatial manifold and taking the large-$N$ limit it is possible to still have a Hagedorn temperature and a deconfinement transition, even if the gauge theory is not confining on $R^3$ \cite{Witten:1998zw}.  In particular, for the case of planar $\NN=4$  super Yang-Mills on $S^3$, one finds a coupling dependent Hagedorn temperature.   This was first computed at zero coupling by Sundborg \cite{Sundborg:1999ue} and independently in \cite{Aharony:2003sx},
 where they found that $T_H=(2\ln(2+\sqrt{3}))^{-1}$ in units where the radius of the $S^3$ is set to $1$.  Furthermore, for this zero coupling case it was shown that $T_H$ equals  the deconfinement temperature \cite{Sundborg:1999ue,Aharony:2003sx}.    However, at large 't Hooft coupling $\lambda$, where $\lambda=g_{YM}^2N$, one expects the deconfinement temperature to be the same as the Hawking-Page transition temperature $T_{HP}\sim1$ for the supergravity dual \cite{Hawking:1982dh,Witten:1998zw}, while $T_H\sim \lambda^{1/4}$, therefore in this regime $T_H$ is far above the deconfinement transition.  However, it still provides information about the growth of states at large energies, or equivalently, the growth of single-trace operators at large dimension.

 Recently, Harmark and Wilhelm have exploited the integrability of planar $\NN=4$ SYM to make substantial progress in finding the coupling dependence of $T_H$ \cite{Harmark:2017yrv,Harmark:2018red,Harmark:2021qma}.  From the $Y$-system  they computed $T_H$ up to the two-loop level at weak coupling \cite{Harmark:2017yrv}, which  previously was only  known  to one loop \cite{Spradlin:2004pp}.  Switching over to the quantum spectral curve formalism \cite{Gromov:2013pga,Gromov:2014caa,Gromov:2015wca} they extended their results to seven loops at weak coupling and found the asymptotic behavior numerically at strong coupling \cite{Harmark:2018red,Harmark:2021qma}.

 At strong coupling Harmark and Wilhelm found that $T_H$ has the form
 \be\label{HWres}
T_H=\sum_{n=0}^\infty c_n g^{(1-n)/2}\approx 0.3989g^{1/2}+0.159-0.0087 g^{-1/2}+0.037 g^{-1}+\dots\,,
\ee
where  $g\equiv \frac{\sqrt{\lambda}}{4\pi}$ and the uncertainties are of order $1$ in the last digit for $c_0$ and $c_1$ and order $5$ in the last digit for $c_2$ and $c_3$.  The first coefficient matches within the error bars the flat-space prediction $c_0=\frac{1}{\sqrt{2\pi}}\approx 0.39894$, which follows from the original computation of the superstring Hagedorn temperature \cite{Sundborg:1984uk} and the AdS/CFT dictionary ({\it c.f.} \cite{Aharony:2003sx} and references therein).  After \eqref{HWres} in \cite{Harmark:2021qma} was announced, the second term was  derived analytically by Maldacena and Urbach \cite{Maldacena_unpub,Urbach:2022xzw}, where they found that $c_1=\frac{1}{2\pi}\approx0.159155$. 

We will rederive this last result and also conjecture the analytic result for the next two terms in the expansion. To do this we will combine analysis from supergravity, where we take advantage of the lightness of the tachyon mode, as well as properties of the plane-wave limit of $AdS_5\times S^5$.  In fact, we can derive the Maldacena-Urbach result from the multiplicity of states in the plane-wave limit. 
Our conjecture for the expansion in \eqref{HWres} is 
\be\label{THan}
T_H= \frac{1}{\sqrt{2\pi}}g^{1/2}+\frac{1}{2\pi}+\left(\left[\frac{5}{8\pi\sqrt{2\pi}}\right]+\left\{\frac{-4\ln(2)}{4\pi\sqrt{2\pi}}\right\}\right)g^{-1/2}+\left[\frac{45}{128\pi^2}\right] g^{-1}+\dots\,,
\ee
where the terms in the square brackets come from  supergravity corrections while the term in the curly brackets comes from a shift in the string zero-point energy.  

To test the conjecture we take  the QSC prescription described in \cite{Harmark:2021qma} and go to higher orders in the series expansions for the $Q$-functions to improve the numerical estimates.  We find
\be\label{THus}
T_H&\approx& 0.39894g^{1/2}+0.15916-0.00865 g^{-1/2}+0.0356 g^{-1}\nn\\
&&\qquad\qquad\qquad\qquad\qquad\qquad-0.008196 g^{-3/2}-0.00671 g^{-2}+\dots\,,
\ee
where the error is $1$ in the last digit for the first four coefficients and $3$ for the last two.  The coefficient $c_1$ is computed assuming that $c_0$ takes its analytic value, hence improving its error bars, while  $c_2$ and $c_3$ are computed together  assuming that $c_1$ also takes its analytic value.  The results for $c_4$ and $c_5$  assume that $c_2$ and $c_3$ take the form in \eqref{THan}.

The conjectured $c_2$ and $c_3$ are approximately $-0.0086538$ and $0.03562$ respectively, so we see that they are in very good agreement with the numerical approximations in \eqref{THus}.
It turns out that our prediction for $c_3$ is more robust than our prediction for $c_2$.  If we assume that $c_3$ equals its conjectured analytic value in \eqref{THan} then our numerical prediction for $c_2$ improves to $c_2=-0.0086538\pm 3\times 10^{-7}$, which is right on the nose with the conjectured result.  The   zero-point shift that affects $c_2$ is twice as large as the shift found in the plane-wave limit.  At the moment we are unable to derive this shift from first principles.

The rest of this paper is organized as follows.  In section 2 we rederive the Maldacena-Urbach result by equating the supergravity calculation for the Hagedorn temperature to finding the ground-state energy for a four-dimensional harmonic oscillator.  We then use first and second order perturbation theory on the perturbed oscillator to find the corrections in the square brackets in \eqref{THan}.  In section 3 we consider the plane-wave limit for $AdS_5\times S^5$ and show that the growth of states leads to the analytic result for $c_1$.  We also provide two arguments why the plane-wave captures the correct result.  We further show that the shift to the zero-point energy will lead to half the result in the curly brackets in \eqref{THan} while also showing that it does not contribute to $c_3$.  In section 4 we summarize the QSC prescription and give further details about our numerical results.

In a companion paper  \cite{toappear} we study the coupling dependence of the Hagedorn temperature for ABJM theory in the planar limit, where we find similar behavior.

\noindent{\it Note added:}  As this paper was being prepared \cite{Bigazzi:2023oqm} appeared which claims to find the first correction to the Hagedorn temperature for the Witten  D4 black brane background from the type IIA world-sheet, matching the supergravity result in \cite{Urbach:2023npi}.

\section{Corrections from supergravity}
The Hagedorn temperature, $T_H$ in string theory is inversely proportional to the string length.  This means that for $\NN=4$ SYM on $S^3$, $T_H\sim \lambda^{1/4}$, where we have chosen units where the radius of the $S^3$ is unity.  It is also known that $\NN=4$ SYM undergoes a deconfinement transition at the Hawking-Page temperature \cite{Hawking:1982dh,Witten:1998zw}, which at strong coupling is $T_{HP}=\frac{3}{2\pi}$.   This is well below $T_H$, hence at $T=T_H$ there is not actually a transition (at least at strong coupling) \cite{Aharony:2003sx}.  However, $T_H$ still provides information about the density of single trace states at zero temperature.  That is, we expect that $\rho(E)\sim e^{E/T_H}$.   

Since we are really working at $T=0$, the relevant geometry to consider in the gravity dual is {\it not} the AdS black hole.  Instead we should consider empty Euclidean $AdS_5\times S^5$ with the Euclidean time direction  $\tau$ identifiied by  $\tau\equiv \tau+\beta$.  The metric for Euclidean $AdS_5$ in global coordinates can be written as 
\be
ds^2=(1+R^2)d\tau^2+\frac{dR^2}{1+R^2}+R^2 d\Omega_3^2\,,
\ee
where $R\ge0$ and $d\Omega_3^2$ is the unit metric on $S^3$.  The string tension is given by ${2\pi}/
{\al'}=2\pi\sqrt{\lambda}$.

\subsection{Zeroth order}

To find the Hagedorn temperature we consider a winding string that wraps around the $\tau$ direction \cite{Sathiapalan:1986db,Kogan:1987jd,Atick:1988si}.  At the same time the fermions with odd winding states have anti-periodic boundary conditions, hence the world-sheet ground state energy is shifted to $-\frac{2}{\al'}$ in the flat-space limit.  Hence the corresponding mass squared for a single winding state in the flat-space limit is
\be
m^2=\left(\frac{\beta}{2\pi\al'}\right)^2-\frac{2}{\al'}\,.
\ee
The Hagedorn temperature is determined by setting $m^2$ to zero, to which we find
\be
T_H=\frac{1}{\beta_H}=\frac{\lambda^{1/4}}{\sqrt{8\pi^2}}\,.
\ee

To find the next term in the expansion, we note that corrections to the world-sheet ground state energy in the single winding sector should be of order $\OO(\lambda^0)$, and hence will not affect the next term in the expansion.  In fact, since we are assuming that we are tuning the mass to be very small, we can treat the problem as that for a point-particle in a supergravity background.  The winding mode is a scalar field, $\chi$,  hence its contribution to the action is
\be 
\int d^5X \sqrt{g}\left(\nabla^\mu\chi\nabla_\mu\chi+m^2(R)\chi^2\right)\,,
\ee
where $m^2(R)$ is the radial dependent mass term
\be\label{msq}
m^2(R)=(1+R^2)\left(\frac{\beta}{2\pi\al'}\right)^2+C\,,
\ee
with $C\approx -{2}/{\al'}$.  

Assuming that $\chi$ has a nontrivial profile only in the $R$ direction and minimizing the action, we find the equations of motion are
approximately
\be\label{HOrd}
-\frac{1}{R^3}\frac{d}{dR} R^3 \frac{d}{dR} \chi(R)+\left(\frac{\beta}{2\pi\al'}\right)^2R^2\chi(R)=\left(\frac{2}{\al'}-\left(\frac{\beta}{2\pi\al'}\right)^2\right)\chi(R)\,.
\ee
The solution should be normalizable and hence fall off to zero as $R\to\infty$.  Therefore, solving \eqref{HOrd} is equivalent to finding the ground-state solution for a rotationally symmetric four-dimensional harmonic oscillator with $\om=\frac{\beta}{2\pi\al'}$ and energy 
\be\label{eneq}
E=\frac{1}{2}\left(\frac{2}{\al'}-\left(\frac{\beta}{2\pi\al'}\right)^2\right)=2\om=2\frac{\beta}{2\pi\al'}\,,
\ee
which leads to
\be\label{betaeq}
\frac{\beta^2}{4\pi\al'}=2\pi-2\beta\,.
\ee
To leading order, the solution of \eqref{betaeq} gives the Hagedorn temperature
\be\label{TH0}
T_H=\frac{1}{2\pi\sqrt{2\al'}}+\frac{1}{2\pi}+\dots\,.
\ee

\subsection{First order}
At the next order we have to consider first order corrections in $\al'$ to the world-sheet sigma model.  We write
$C=-\frac{2}{\al'}+\Delta C$ to take this correction into account.  There is also the correction to the harmonic oscillator ``Hamiltonian",
\be
\Delta H=-\frac{1}{2}\frac{1}{R^3}\frac{d}{dR} R^5 \frac{d}{dR}\,,
\ee
which leads to the correction to the energy
\be
\Delta E^{(1)}=\langle\psi_0|\Delta H|\psi_0\rangle=3\,,
\ee
where 
 the normalized\footnote{We set $\displaystyle\int_0^\infty R^3\psi_0(R)\psi_0(R) dR=1$.} ground-state wave-function is $\psi_0(R)=\sqrt{2}\om e^{-\frac{1}{2}\om R^2}$, we find
\be
\langle\psi_0|\Delta H^{(1)}|\psi_0\rangle=\langle\psi_0|\frac{1}{2}\left(2\om R^2-\om^2\R^4\right)|\psi_0\rangle=-1\,.
\ee
Hence we find,
\be\label{eneq1}
2\frac{\beta}{2\pi\al'}+3=\frac{1}{2}\left(\frac{2}{\al'}-\Delta C-\left(\frac{\beta}{2\pi\al'}\right)^2\right)\,,
\ee
which we rewrite as
\be\label{betaeq1}
\frac{\beta^2}{4\pi\al'}=2\pi-2\beta-{\pi\al'}\Delta C-6\pi\al'\,.
\ee
We will later conjecture that $\Delta C$ has the form $\Delta C=\frac{\beta^2}{2\pi^2\alpha'}\Delta c$, in which case solving for 
 $T_H$ to the next order we find
\be\label{TH1}
T_H&=&\frac{1}{2\pi\sqrt{2\al'}}+\frac{1}{2\pi}+\frac{5+2\Delta c}{4\sqrt{2}\pi}\sqrt{\al'}+\dots\nn\\
&=&\frac{\sqrt{g}}{\sqrt{2\pi}}+\frac{1}{2\pi}+\frac{5+2\Delta c}{8\pi\sqrt{2\pi}}\frac{1}{\sqrt{g}}+\dots\,,
\ee
where $g$ is defined as $g=\frac{\sqrt{\lambda}}{4\pi}=\frac{1}{4\pi\al'}$.

\subsection{Second order}

If we assume that the corrections to $C$ are in even powers of $\beta$ only,  then the next order correction to $\Delta C$ won't affect the second order result.  The correction to the energy is computed using second order perturbation theory, where we find
\be\label{E2}
\Delta E^{(2)}&=&\frac{\langle\psi_0|\Delta H^{(1)}|\psi_2\rangle\langle\psi_2|\Delta H^{(1)}|\psi_0\rangle}{-2\,\om}+\frac{\langle\psi_0|\Delta H^{(1)}|\psi_4\rangle\langle\psi_4|\Delta H^{(1)}|\psi_0\rangle}{-4\,\om}\nn\\
&=&0+\frac{(-\sqrt{3})^2}{-4\,\om}=-\frac{3\pi\al'}{2\,\beta}\,.
\ee
In computing \eqref{E2} we used that the excited normalized wave-functions are given by
\be
\psi_2(R)=\om(\om R^2-2)e^{-\frac{1}{2}\om R^2}\,,\qquad \psi_4(R)=\frac{\om}{\sqrt{6}}(\om^2 R^4-6\om R^2+6)e^{-\frac{1}{2}\om R^2}\,.
\ee
Hence we reach the equation
\be\label{eneq2}
2\frac{\beta}{2\pi\al'}+3-\frac{3\pi\al'}{2\,\beta}=\frac{1}{2}\left(\frac{2}{\al'}-\frac{\beta^2\Delta c}{2\pi^2\al'}-\left(\frac{\beta}{2\pi\al'}\right)^2\right)\,,
\ee
which is equivalent to
\be\label{betaeq2}
\frac{\beta^2}{4\pi\al'}=2\pi-2\beta-\frac{\beta^2\Delta c}{2\pi}-6\pi\al'+\frac{3\pi^2(\al')^2}{\beta}\,.
\ee
Hence the Hagedorn temperature as a function of $g$ is
\be\label{TH3}
T_H(g)&=&\frac{\sqrt{g}}{\sqrt{2\pi}}+\frac{1}{2\pi}+\frac{5+2\Delta c}{8\pi\sqrt{2\pi}\sqrt{g}}+\frac{45}{128\pi^2g}+\OO(g^{-3/2})\nn\\
&=&0.398942\sqrt{g}+0.159155+\frac{0.079367+0.0317468\,\Delta c }{\sqrt{g}}+\frac{0.0356207}{g}+\OO(g^{-3/2})\,.\nn\\
\ee
Notice that the last term is independent of $\Delta c$.


\section{The Hagedorn temperature from the plane-wave}

As we have indicated, the Hagedorn temperature is dependent on the growth of string states as the energy increases.  Unfortunately, the exact spectrum is unknown for type IIB string theory on $AdS_5\times S_5$.  However, it is instructive to consider the IIB theory in the plane-wave limit where the spectrum is known.

The analysis of the Hagedorn temperature in the plane-wave limit was originally carried out in
 \cite{PandoZayas:2002hh,Greene:2002cd,Grignani:2003cs} where an exact equation  for the Hagedorn temperature in the plane wave background was derived.  Here we will review this analysis, mainly following the discussion in \cite{Grignani:2003cs}.  We then show that in the plane-wave limit, one finds the same first order correction for the Hagedorn temperature found numerically in \cite{Harmark:2021qma} and using the supergravity dual in \cite{Maldacena_unpub,Urbach:2022xzw}.  We further show that the zero-point energy leads to a correction to the second-order term in the expansion, but not the third-order term.

Consider the full metric for the $AdS_5\times S^5$ type IIB background,
\be\label{ppmetric1}
ds^2&=&-(1+R^2)dt^2+\frac{dR^2}{1+R^2}+R^2 d\Omega_3^2+(1-Z^2)d\psi^2+\frac{dZ^2}{1-Z^2}+Z^2d{\Omega_3'}^2\nn\\
&=&\bigg(-(1+R_I^2)dt^2+(1-Z_I^2)d\psi^2+dR_I^2+dZ_I^2\bigg)+
\left\{-\frac{R^2dR^2}{1+R^2}+\frac{Z^2dZ^2}{1-Z^2}\right\}\\
\label{ppmetric2}
&=&\bigg(-2dx^+dx^--\frac{1}{2}(R_I^2+Z_I^2)dx^+dx^++dR_I^2+dZ_I^2\nn\\
&&-\left[({R}^2-Z^2) dx^+dx^-+\frac{1}{2}({R}^2+{Z}^2)dx^-dx^-\right]\bigg)+
\left\{-\frac{R^2dR^2}{1+R^2}+\frac{Z^2dZ^2}{1-Z^2}\right\},
\ee
where $x^{\pm}=\frac{1}{\sqrt{2}}(t\pm\psi)$ and  $I=1,\dots 4$.  If we consider geodesics where $R^2,\ Z^2\ll1$, then we can drop the term in the curly brackets.  If we further focus on geodesics with large angular momentum along $\psi$, then $\dot x^-\ll\dot x^+$ and we can drop the term in the square brackets, leaving the plane-wave metric
\be\label{ppwave}
ds^2=-2dx^+dx^--f^2x_I^2dx^+dx^++dx_I^2\,,
\ee
 where $I=1,\dots 8$ and  $f=1/\sqrt{2}$.  
There is also a background Ramond-Ramond five-form field-strength,
\be\label{fs5}
F_{+1234}=F_{+5678}=2f\,.
\ee
In terms of the dimension $\Delta$ and $R$-charge $J$ of the dual operators, we have
\be
p^{-}=\frac{f}{\sqrt{2}}(\Delta-J)\,,\qquad\qquad p^{+}=\frac{\Delta+J}{\sqrt{2}f}\,.
\ee

Type IIB string theory on the plane-wave metric in \eqref{ppwave} with the  field-strength in \eqref{fs5} is exactly solvable and can be quantized in light-cone gauge \cite{Metsaev:2001bj,Metsaev:2002re}. In particular, one finds for the light-cone Hamiltonian
\be
H=P^{-}=f(N_0^B+N_0^F+4)+\frac{1}{\al' p^+}\sum_{{\II}=1}^2\sum_{m=1}^\infty\sqrt{m^2+(\al'p^+f)^2}(N_{\II m}^B+N_{\II m}^F)\,,
\ee
where $\II$ counts the fermionic world-sheet variables, $N_m^B$ counts the bosonic oscillators at level $m$ and $N_m^F$ counts the fermionic oscillators at level $m$.  More details can be found in \cite{Metsaev:2002re}.

The free energy is the sum over the free energies for each species of the single string spectrum.  The bosons contribute
\be
\frac{1}{\beta}\Tr\ln\left(1-e^{-\beta p^0}\right)=-\sum_{n=1}^\infty\frac{1}{n\beta}\Tr\, e^{-\frac{n\beta}{\sqrt{2}}(p^++p^{-})}\,,
\ee
while the fermions contribute
\be
-\frac{1}{\beta}\Tr\ln\left(1+e^{-\beta p^0}\right)=\sum_{n=1}^\infty\frac{(-1)^n}{n\beta}\Tr\, e^{-\frac{n\beta}{\sqrt{2}}(p^++p^{-})}\,.
\ee
The trace over states includes an integration over $p^{+}$ which is normalized to \cite{Grignani:2003cs}
\be
\frac{L}{\pi\sqrt{2}}\int_0^\infty dp^+\,,
\ee
where $L$ is the length along the spatial part of the light-cone.  If we change variables to $\tau_2$, where
\be\label{tau2}
\tau_2=\frac{n\beta}{2\pi\sqrt{2}\al'p^+}, 
\ee
then averaging over the free energy from either bosons and fermions leads to
\be
F=-\sum_{n=1,\tt odd}^\infty\frac{L}{4\pi^2\al'}\int_0^\infty \frac{d\tau_2}{\tau_2^2}e^{-\frac{n^2\beta^2}{4\pi\al'\tau_2}}\tr \, e^{-\frac{n\beta}{\sqrt{2}}p^{-}}-\frac{L\pi}{24\beta^2}\,,
\ee
where $\tr$ refers to the trace over all possible string oscillator modes.   The last term arises because of a mismatch between the number of bosonic and fermionic states at $p^{-}=0$.  Explicitly writing the oscillators and enforcing level matching by introducing an integral over $\tau_1$, the free energy becomes
\be
F=-\sum_{n=1,\tt odd}^\infty\frac{L}{4\pi^2\al'}\int_0^\infty \frac{d\tau_2}{\tau_2^2}\int_{-\frac{1}{2}}^{\frac{1}{2}}d\tau_1e^{-\frac{n^2\beta^2}{4\pi\al'\tau_2}}\prod_{m=-\infty}^\infty\left(\frac{1+e^{-2\pi\tau_2\sqrt{m^2+\nu^2}+2\pi i\tau_1 m}}{1-e^{-2\pi\tau_2\sqrt{m^2+\nu^2}+2\pi i\tau_1 m}}\right)^8-\frac{L\pi}{24\beta^2}\,,\nn\\
\ee
where $\nu=\frac{n\beta f}{2\pi\sqrt{2}\tau_2}$.  We then observe that 
\be\label{prfor}
\prod_{m=-\infty}^\infty\left(\frac{1+e^{-2\pi\tau_2\sqrt{m^2+\nu^2}}}{1-e^{-2\pi\tau_2\sqrt{m^2+\nu^2}}}\right)^8&=&
\exp\left(-8\sum_{m=-\infty}^\infty\sum_{p=1}^\infty\left[\frac{(-1)^p}{p}-\frac{1}{p}\right]e^{-2\pi p\tau_2\sqrt{m^2+\nu^2}}\right)\nn\\
&=&\exp\left(2\sum_{m=-\infty}^\infty\sum_{p=1, \tt odd}^\infty \frac{1}{p}\frac{16}{\sqrt{\pi}}\int_0^\infty dt\, e^{-t^2-\frac{\pi^2\tau_2^2p^2m^2}{t^2}-\frac{n^2\beta^2f^2p^2}{8t^2}}\right)\nn\\
&\approx&\exp\left(\frac{8n\beta f}{\pi\sqrt{2}\,\tau_2}\sum_{p=1}^\infty\frac{1-(-1)^p}{p}K_1\left(\frac{n\beta f p}{\sqrt{2}}\right)\right)\,,
\ee
where $K_1(x)$ is the modified Bessel function. To get to the last step we approximated the sum over $m$ as an integral, since in the following step we will assume that $\tau_2\to0$.

The Hagedorn temperature is determined by taking $\tau_1,\tau_2\to 0$.  Using the results from above, the free energy behaves as
\be\label{Fhag}
F\sim -\sum_{n=1,\tt odd}^\infty\frac{L}{4\pi^2\al'}\int_0^\infty \frac{d\tau_2}{\tau_2^2}e^{-\frac{n^2\beta^2}{4\pi\al'\tau_2}}\exp\left(\frac{8n\beta f}{\pi\sqrt{2}\,\tau_2}\sum_{p=1}^\infty\frac{1-(-1)^p}{p}K_1\left(\frac{n\beta f p}{\sqrt{2}}\right)\right)\,.
\ee
The dominant contribution comes from $n=1$, which corresponds to a single winding around the thermal circle.  The free energy diverges as $\tau_2\to0$ when the relation
\be
\frac{\beta^2}{4\pi\al'}=\frac{8\beta f}{\pi\sqrt{2}}\sum_{p=1}^\infty\frac{1-(-1)^p}{p}K_1\left(\frac{\beta f p}{\sqrt{2}}\right)\,
\ee
is satisfied.
Solving this equation for $\beta$ determines the Hagedorn temperature for the plane-wave.  This equation can be expressed as a series in $\beta$ by implementing a Mellin transform on the right hand side and then inverting the transform.  After these steps one finds 
\be\label{hageq}
\frac{\beta^2}{4\pi\al'}&=&2\pi-{2\beta (\sqrt{2}f)}+\frac{\beta^2(\sqrt{2}f)^22\ln2}{2\pi}\nn\\
&&\qquad\qquad -\sum_{k=2}^\infty\frac{(-1)^k(2^{2k}-4)4\sqrt{\pi}}{k!}\left(\frac{\beta (\sqrt{2}f)}{4\pi}\right)^{2k}\Gamma(k-\frac{1}{2})\zeta(2k-1)\,.
\ee

If we now set $f=1/\sqrt{2}$ to match the plane-wave limit of $AdS_5\times S^5$  and compare \eqref{hageq} to \eqref{betaeq2}, we see that the linear term in $\beta$ is the same in both equations. Hence, the first order correction to the Hagedorn temperature matches the Maldacena-Urbach result from supergravity.    Given the success  matching this term,  it is tempting to set $\Delta c=-2\ln2$ in \eqref{betaeq2} so that the terms quadratic in $
\beta$ also match.  The higher order terms in \eqref{hageq} lead to corrections of order $g^{-3/2}$ and higher and are irrelevant for this study, while the last two terms in \eqref{betaeq2} are from non-quadratic corrections.

Note that the right hand side of \eqref{hageq} with the second term removed is an even function of $\beta$.  The second term is the correction coming from the winding mode on the thermal circle with the extra quadratic piece.  The even terms make up the zero-point energy on the string world-sheet.  The first term is the flat-space contribution, while the third term is the leading order correction.

We now give two arguments why the plane-wave leads to the correct result for the leading order correction to the Hagedorn temperature.  In this analysis  the Hagedorn temperature is found by taking $\tau_2\to0$ in \eqref{Fhag}, such that $\tau_2\ll \frac{\beta^2}{\al'}\sim1$.  Using \eqref{tau2} this corresponds to having $p^+\gg \beta^{-1}$.   On the other hand, the modes that contribute to the second line of \eqref{prfor} cut off around $m\sim 1/\tau_2$, hence these have $p^{-}\sim \beta^{-1}$.  Therefore, the regime that determines the Hagedorn temperature has $p^{+}\gg p^{-}$, which focuses on the plane-wave limit.  This suggests that including the term inside the square brackets in  \eqref{ppmetric2} would not affect the Hagedorn temperature, at least to leading order, since as $\tau_2\to0$ the contribution to the free energy would approach that in  \eqref{Fhag}.  But this term plus the plane-wave metric  gives the quadratic metric inside the parentheses in \eqref{ppmetric1}.  Since it is effectively this metric that was used in \cite{Maldacena_unpub,Urbach:2022xzw} to compute the leading order correction to the Hagedorn temperature,  this shows why  the plane-wave limit can give the correct result.

A second way to see why the plane-wave limit can give the correct result is by turning on a chemical potential $\mu$ for the charge $J=\frac{1}{\sqrt{2}}(p^+-p^-)$.  The effect of the chemical potential is to modify the free energy to \cite{Greene:2002cd,Grignani:2003cs}
\be
F\sim \Tr\ln\left(-\partial_+\partial_-+{\frac{\beta^2-\mu^2}{4\pi\al'}-\left[2\pi-2(\beta-\mu)+\frac{(\beta-\mu)^2\ln2}{\pi}+\dots\right]}\right)\,.
\ee
Hence, the expectation value of  $J$ is
\be
\langle J\rangle =\beta\frac{\partial F}{\partial\mu}\sim \Tr \frac{-\frac{\mu}{2\pi\al'}-2+\dots}{\left(-\partial_+\partial_-+\frac{\beta^2-\mu^2}{4\pi\al'}-\left[2\pi-2(\beta-\mu)+\frac{(\beta-\mu)^2\ln2}{\pi}+\dots\right]\right)}\,.
\ee
Clearly, $\langle J\rangle<0$ if $\mu=0$.  This nonzero value for $\langle J\rangle$ is due to the non-symmetric nature of the plane-wave metric.  To compensate for this  we  set $\mu=-4\pi\al'+\dots$.  This results in the new Hagedorn equation
\be\label{chemhag}
\frac{\beta^2}{4\pi\al'}=2\pi-2\beta+\frac{\beta^2\ln2}{\pi}-4\pi\al'+\dots\,.
\ee
Since the term linear in $\beta$ is unchanged, the first  correction to the Hagedorn temperature also remains unchanged.  However, the last term in \eqref{chemhag} will affect the next order correction.

\section{Numerical results}

To test our predictions we carry out and extend the QSC analysis of Harmark and Wilhelm in \cite{Harmark:2021qma}.   To determine the asymptotic behavior they computed $T_H$ from $0\le\sqrt{g}\le1.8$ in steps of $\Delta\sqrt{g}=0.025$.  Their results for $T_H$ are accurate to six significant digits for $\sqrt{g}=1.8$ and improve significantly for smaller values of $\sqrt{g}$.  
In our analysis we are able to push $\sqrt{g}$ out to $2.25$ where we conservatively estimate that the result for $T_H$ is accurate to at least $1\times10^{-9}$.  This allows us to make more precise estimates for the asymptotic coefficients.

  We briefly sketch the QSC procedure of Harmark and Wilhelm.  More details can be found in \cite{Harmark:2021qma} \footnote{We keep the same convention as in \cite{Harmark:2021qma}, which reverses the  $spatial$ and $R$-symmetry indices in the $Q$-functions.}.  The starting point are the 256 $PSU(2,2|4)$ $Q$-functions which satisfy a set of difference equations.  The ones most important for our purposes are
\be\label{Qeq}
Q_{a|i}^{+}(u)-Q_{a|i}^{-}(u)+\bQ_i(u)\bQ^j(u)Q_{a|j}^{+}(u)=0\,,
\ee
where $a=\{1,2,3,4\}$, $i=\{1,2,3,4\}$, $Q_{a|i}^{\pm}(u)=Q_{a|i}(u\pm \frac{i}{2})$, and $\bQ^i(u)=\chi^{ij}\bQ_j(u)$, with
\be
\chi^{ij}=\left(\begin{array}{cccc}
    0 & 0&0&-1 \\
    0 & 0&+1&0 \\
     0 &-1& 0&0 \\
     +1&0 & 0&0
\end{array}
\right)\,.
\ee
We have other $Q$-functions, $\bP_a(u)$, which satisfy the relation 
\be\label{PQeq}
\bP_a(u)=-\bQ^i(u)Q_{a|i}^+(u)\,.
\ee
In addition, there are also orthogonality conditions
\be\label{ortho}
Q_{a|i}Q^{b|i}=-\delta_a^b\,,\qquad Q_{a|i}Q^{a|j}=-\delta_i^j\,,
\ee
where $Q^{a|i}=\chi^{ab}\chi^{ij}Q_{b|j}$.

One must also impose asymptotic and analytic conditions for the $Q$-functions.  The asymptotic conditions follow from the $T$-system and have the behavior
\be
\bP_1(u)&=&A_1y^{-i u}\,,\nn\\
\bP_2(u)&=&A_2\left(u-i \frac{1-4y^2+y^4}{2(1-y^4)}\right)y^{-i u}\,,\nn\\
\bP_3(u)&=&A_3y^{i u}\,,\nn\\
\bP_4(u)&=&A_4\left(u+i \frac{1-4y^2+y^4}{2(1-y^4)}\right)y^{i u}\,,\\
\bQ_1(u)&=&B_1\,,\qquad \bQ_2(u)=B_2 u\,,\qquad \bQ_3(u)=B_3 u^2\,,\nn\\
\bQ_4(u)&=&B_4u\left(u^2+\frac{5-2y^2+5y^4}{(1+y^2)^2}\right)\,,\nn
\ee
where $y=e^{i\pi}\exp(-1/(2T_H))$.  Using the so-called $H$-symmetry, the coefficients can be fixed to 
\be
&&A_1=iA_2=-A_3=-iA_4=\frac{1-y}{1+y}\nn\\
&&B_1=B_2=1\,,\qquad B_3=-\frac{i}{2(1-y)^4}\,,\qquad B_4=-\frac{i}{6(1-y)^4}\,.
\ee

The $\bQ_i(u)$ have a single short cut in the $u$ plane between $-2g$ and $2g$.  Hence, we can write the $\bQ_i(u)$ as the series expansions
\be\label{Qexp}
\bQ_1(u)&=&1+\sum_{n=1}^\infty\frac{c_{1,2n}(g)g^{2n}}{x(u)^{2n}}\,,\nn\\
\bQ_2(u)&=&gx(u)\left(1+\sum_{n=1}^\infty\frac{c_{2,2n-1}(g)g^{2(n-1)}}{x(u)^{2n}}\right)\,,\nn\\
\bQ_3(u)&=&-\frac{i}{2(1-y)^4}(gx(u))^2\left(1+\sum_{n=2}^\infty\frac{c_{3,2n-2}(g)g^{2(n-2)}}{x(u)^{2n}}\right)\,,\nn\\
\bQ_4(u)&=&-\frac{i}{6(1-y)^4}(gx(u))^3\left(1+\frac{c_{4,-1}(g) g^{-2}}{x(u)^2}+\sum_{n=2}^\infty\frac{c_{3,2n-3}(g)g^{2(n-2)}}{x(u)^{2n}}\right)\,,
\ee
where the coefficients $c_{i,m}(g)$ are real and $x(u)$ is the Zhukovsky variable,
\be
x(u)=\frac{u}{2g}\left(1+\sqrt{1-\frac{4g^2}{u^2}}\right)\,.
\ee
The $\bP_a$ have a single long-cut  on $u\in \{-\infty,-2g\}$ and $u\in \{+2g,+\infty\}$.  Continuing across the cut we obtain the new function $\widetilde \bP_a(u)$ which satisfies the gluing condition \cite{Gromov:2015wca}
\be\label{glue}
\widetilde \bP_a(u)=(-1)^{1+a}\overline{\bP_a(u)}\,, \qquad u\in\{-2g,2g\}\,.
\ee
The $\widetilde \bP_a(u)$ can be extracted from \eqref{PQeq} by replacing $x(u)$ in \eqref{Qexp} with $\widetilde x(u)=1/x(u)$.

To solve numerically, we  assume that the $Q_{a|i}(u)$ have the large $u$ expansion
\be\label{Qaiexp}
Q_{a|i}(u)=y^{-s_a i u}u^{p_{a|i}}\sum_{n=0}^N \frac{B_{a|i,n}}{u^n}\,,
\ee
where
\be
s_a=1\,,\quad a=1,2\,,\qquad&&\qquad s_a=-1\,,\quad a=3,4\nn\\
p_{a|i}&=&s_a+a+i-3\,,
\ee
and $N$ is some large cutoff.
To match the asymptotics we set the leading coefficients to
\be
B_{a|i,0}=- s_a\frac{\sqrt{y}}{1-y} A_a B_i\,.
\ee
A remaining gauge freedom also allows us to set 
\be
B_{3|i,n}=(-1)^{n+1}B_{1|i,n}\,,\qquad B_{4|i,n}=(-1)^{n+1}B_{2|i,n}\,.
\ee

We then take the difference equation in \eqref{Qeq} and multiply it by a factor to put it in the form
\be\label{Qeq2}
y^{i s_a u}u^{-p_{a|i}}\left(Q_{a|i}^+-Q_{a|i}^-+\bQ_i\bQ^jQ_{a|j}^+\right)=\sum_{n=1}^\infty u^{3-n}V_{a|i,n}\,,
\ee
where the $V_{a|i,n}=0$.  Solving these in order leads to a set of linear equations for the $B_{a|i,n}$.  Solving up to $n=10$ and using the orthogonality relations in \eqref{ortho} then fixes $B_{a|j,m}$ up to $m=2$ in terms of $g$, $y$ and all $c_{i,\ell}$ that appear in the sums in \eqref{Qaiexp} with each sum cut off at $n=K=2$.  Furthermore, solving up to $n=10$ also solves for $B_{a|4,m}$, $B_{a|3,m-2}$ and $B_{a|2,m-2}$ up to $m=10$ in terms of the other coefficients.  One also finds that consistency requires the relations 
\be
c_{4,-1}&=&\frac{1+10y+y^2}{(1+y^2)}+9g^2 c_{2,1}-5g^4 c_{1,2}\\
c_{4,1}&=&-2\frac{1+y+y^2}{(1+y^2)}+3\frac{1+4y+y^2}{(1+y^2)}c_{2,1}-2g^2\frac{1-5y+y^2}{(1+y^2)}c_{1,2}+9g^2c_{3,2}\nn\\
&&\qquad\qquad\qquad-15g^4c_{2,3}+27g^4c_{2,1}c_{1,2}+7g^6c_{1,4}-15g^6 (c_{1,2})^2\,.
\ee
Going beyond $n=10$ to $n=N+8$ then fixes all $B_{a|i,m}$ up to $m=N$ in terms of $g$, $y$ and all $c_{j,\ell}$ up to the cutoff $K=N/2+1$, assuming that $N$ is even.  For a given $g$ this then gives $4K-3$ independent $c_{i,\ell}$ parameters, plus the $y$ parameter.

To fix these $4K-2$ parameters we impose the gluing conditions at the points $u=x_n\equiv 2g \cos( \pi (n-1/2)/I_p)$, $n=1,2,\dots I_P$.  Since the $\bP_a(u)$ are complex on the interval, each point gives two conditions.  However, because of symmetry the point at $x_n$ gives the same conditions as the point at $x_{I_P-n}=-x_n$.  Hence, in order to fix the parameters, the number of points must satisfy $I_P\ge 4K-2$.  To determine $\bP_a(x_n)$ and $\tilde \bP_a(x_n)$ we consider the large $u$ expansion for $Q_{a|i}(u)$ at $u=x_n+i\, U/2$, where $U$ is a large positive odd integer, and use the QSC equations in \eqref{Qeq} to bring $Q_{a|i}(u)$ to the points $u=x_n+i/2$.  From there we can use \eqref{PQeq} to find $P_a(x_n)$ and the corresponding equation to find $\tilde P_a(x_n)$.

\begin{figure}[!tbp]
        \centering
        {\includegraphics[height=0.5\linewidth]{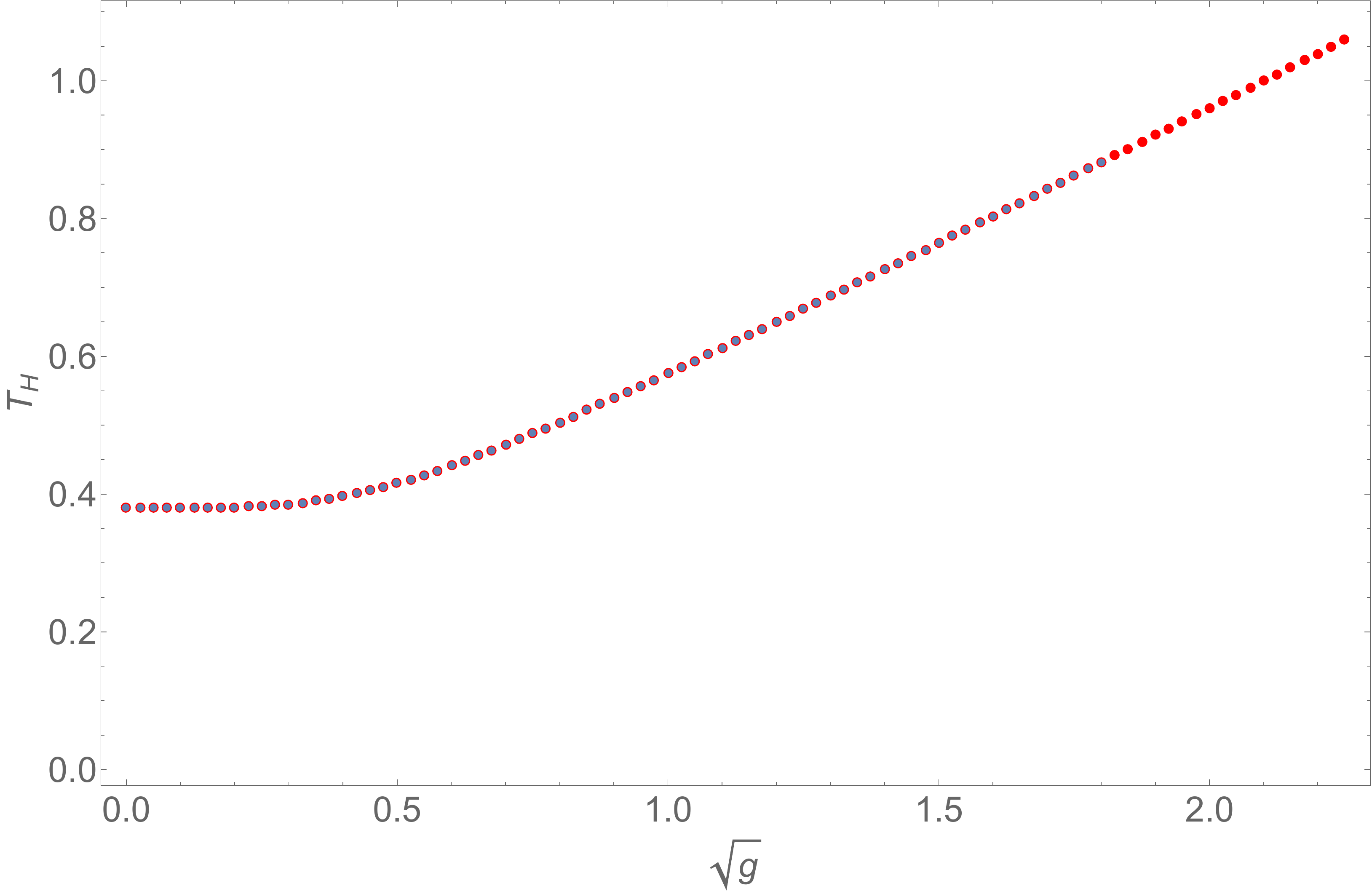}}
                \caption{A plot of $T_H$ versus $\sqrt{g}$.  The results from \cite{Harmark:2021qma} are in blue while our results are in red and lie beneath the blue dots for $\sqrt{g}\le 1.80$.}
        \label{pic:sqrtg_plot}
\end{figure}

For the ranges of $\sqrt{g}$ carried out in our QSC computations we set $N$, $K$, $U$ and $I_P$ to the values shown in table \ref{tab:parms}\,.  The resulting values for $T_H$ are shown in figure \ref{pic:sqrtg_plot}, along with the results found in \cite{Harmark:2021qma}.  The  values computed here are provided in the supplemental file ``tempdata.csv".

\begin{table}[!btp]
    \centering
    \begin{tabular}{|c|c|c|c|c|}
    \hline
   Range of $\sqrt{g}$ &$N$&$K$&$U$&$I_P$\\
     \hline\hline
     $0$ to $1.00$   & $18$ & $10$ &$91$ & $44$ \\
     \hline
      $1.025$ to $1.225$  & $24$ & $13$ &$91$ & $56$\\
      \hline
     $1.25$ to $1.45$ & $30$ & $16$ &$91$ & $68$\\
      \hline
     $1.475$ to $1.65$& $36$ & $19$ &$91$ & $80$\\
      \hline
     $1.675$ to $1.80$ & $42$ & $22$ &$101$ & $104$\\
      \hline
     $1.825$ to $2.05$& $48$& $25$ &$101$ & $124$\\
      \hline
    
     $2.075$ to $ 2.20$& $54$& $28$ &$121$ & $136$\\
      \hline
     $2.225$ to $ 2.25$& $60$& $31$ &$121$ & $148$\\
      \hline
 \end{tabular}
    \caption{Values of $N$, $K$, $U$ and $I_P$ for ranges of $\sqrt{g}$}
    \label{tab:parms}
\end{table}
We assume that the analytic behavior of $T_H$ has the form
\be\label{xcurve}
T_H=c_0 g^{1/2}+c_1+c_2g^{-1/2}+c_3 g^{-1}+c_4 g^{-3/2}+c_5 g^{-2}+\dots.
\ee
To find the coefficients $c_n$ we go in steps, taking advantage of the fact that the first two coefficients are known  analytically. Starting with the leading term, we fit the data by looking for a stable starting point that is as small as possible but is still in the asymptotic regime.  Typically this occurs somewhere between  $0.75<\sqrt{g}<1.0$.  We then take the data points  out to $\sqrt{g}=2.25$ and fit to a curve of the form in \eqref{xcurve} with highest power $g^{-7/2}$.  We find that $c_0=.39894\pm 1\times10^{-5}$,  where the estimated error is mainly due to the uncertainty of the starting point for the fit.  This matches the analytic result $c_0=\frac{1}{\sqrt{2\pi}}\approx 0.39894$. 
Subtracting off $\frac{1}{\sqrt{2\pi}}g^{1/2}$ from the curve, the leading term is now the constant.  Proceeding as before, we  fit this up to $g^{-4}$, where we  find that $c_1=0.15916\pm 1\times 10^{-5}$. This matches nicely with the analytic result $c_1=\frac{1}{2\pi}\approx 0.159155$.

\begin{figure}[!btp]
        \centering
        {\includegraphics[height=0.55\linewidth]{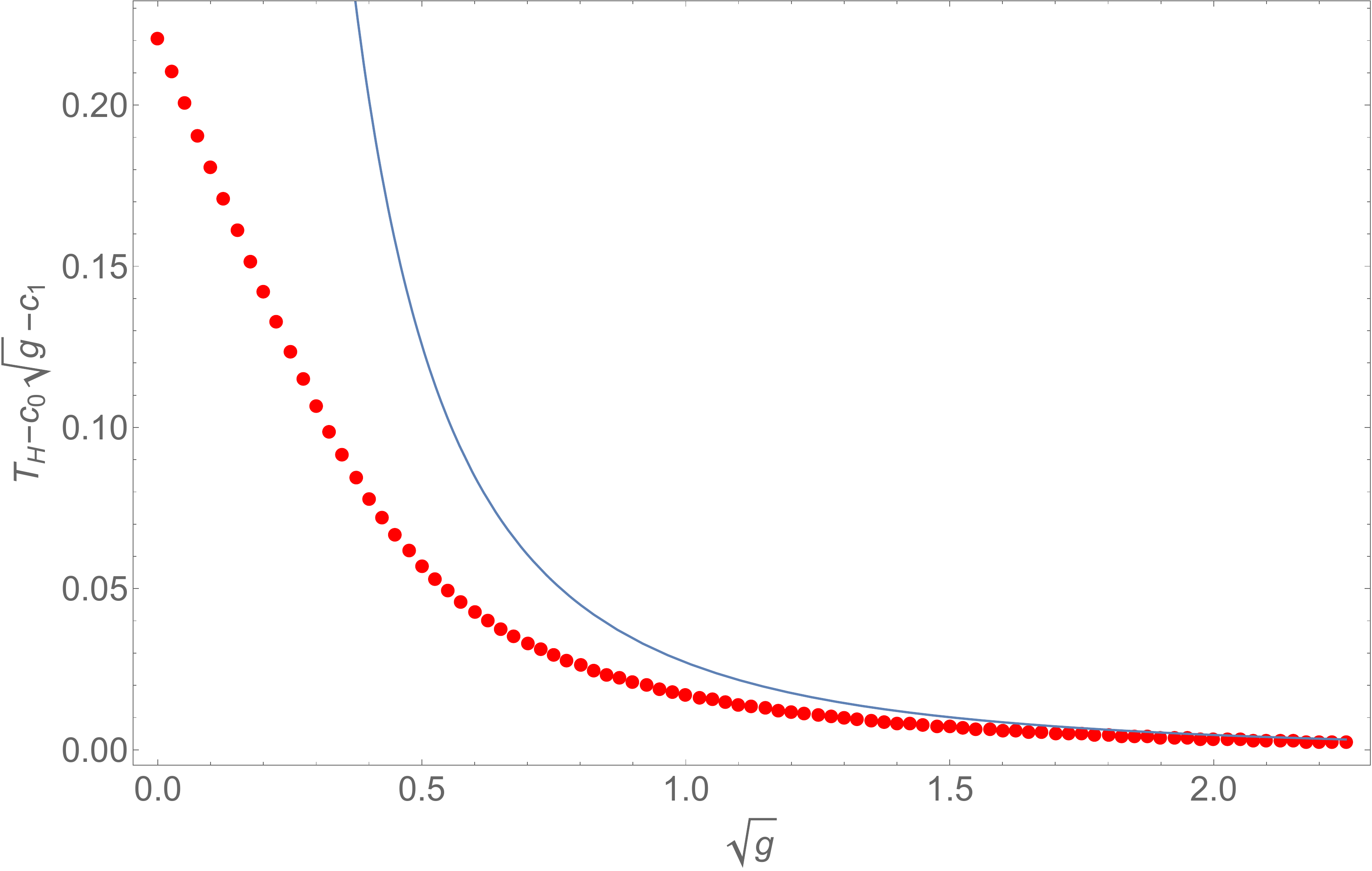}}
                \caption{Plot of $T_H-\frac{\sqrt{g}}{\sqrt{2\pi}}-\frac{1}{2\pi}$ versus $\sqrt{g}$. The plot also shows the curve $c_2 g^{-1/2}+c_3 g^{-1}$ }
        \label{pic:sqrtg_plotmc0c1}
\end{figure}

Next, we subtract off the  analytic result for the constant and are left with the curve shown in figure \ref{pic:sqrtg_plotmc0c1}.  We can then compute the next two terms in the series by fitting to a curve up to $g^{-9/2}$.  We  find that $c_2=-0.00865\pm 1\times 10^{-5}$ and $c_3=0.0356\pm 1\times 10^{-4}$, which improves on the precision of these same two coefficients as found by Harmark and Wilhelm \cite{Harmark:2021qma}.    The error bars on $c_2$ and $c_3$ are correlated.  Moreover, $c_2$ turns out to be unnaturally small, so the $c_3$ term dominates over the range of $\sqrt{g}$ we are able to consider.  With this in mind we analyze the second term first.   If we compare it to the corresponding term in \eqref{TH3}, we see that it matches to three significant digits.  If we now assume that $c_3$ takes the analytic form in \eqref{TH3},  we find that $c_2=-0.0086538\pm 3\times 10^{-7}$ with substantially narrower error bars.    If we borrow from the plane-wave result, then $\Delta c=-
2\ln(2)$, which gives $c_2=0.035366$.  This obviously fails to match the numerical result when plugging into \eqref{TH3}.  However, if we instead choose $\Delta c=-4\ln(2)$, then $c_2=-0.0086538$ which is right on top of our numerical result!  In a companion paper we will show that the corresponding result for the Hagedorn temperature in ABJM theory is consistent with $\Delta c=-3\ln(2)$ \cite{toappear}, suggesting that the coefficient in front of the $\ln(2)$ is related to the dimension of the CFT.

\begin{figure}[!btp]
        \centering
        {\includegraphics[height=0.55\linewidth]{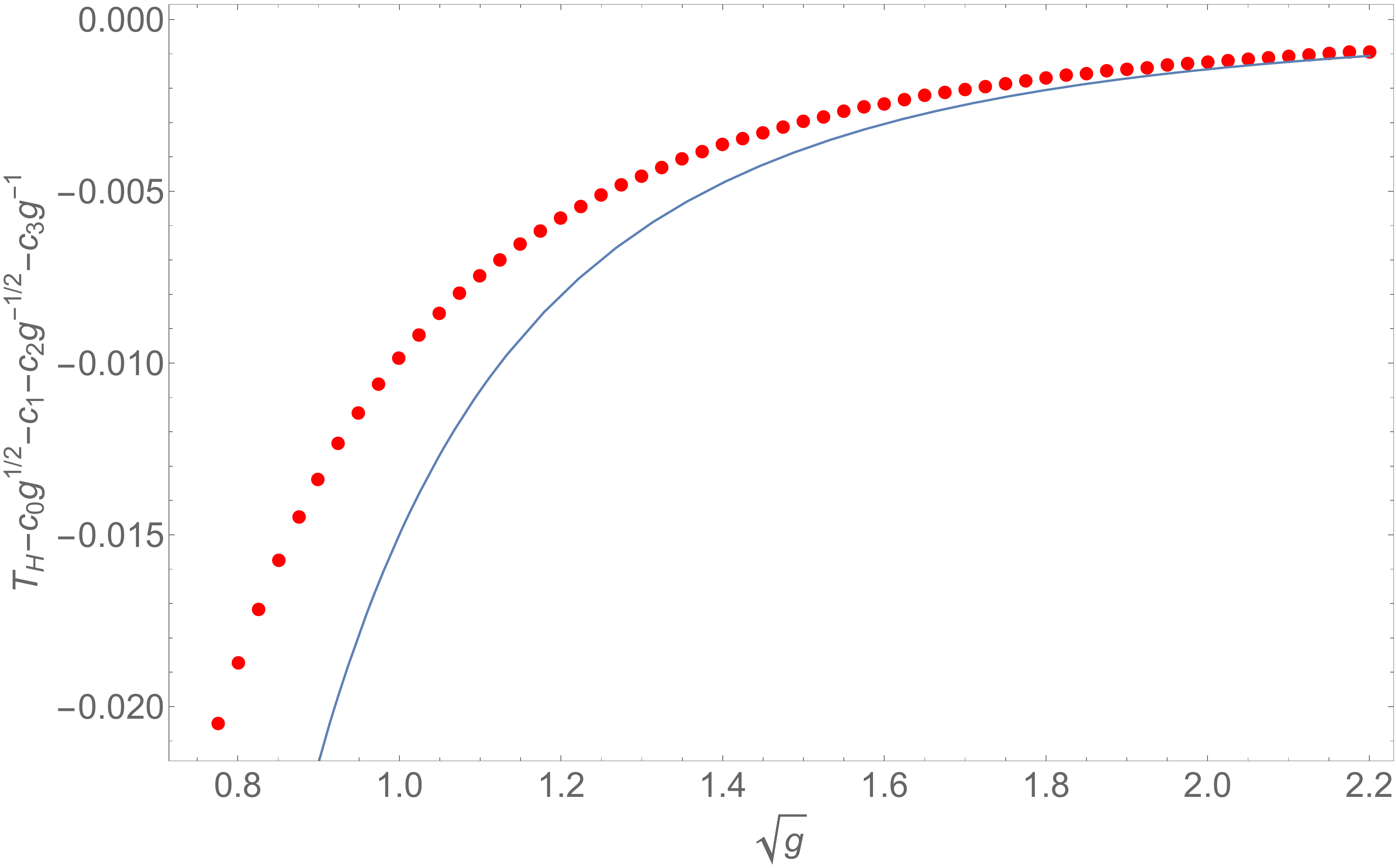}}
                \caption{Plot of $T_H$ versus $\sqrt{g}$ with the first four terms in the expansion subtracted off.  The plot also shows the curve $c_4 g^{-3/2}+c_5 g^{-2}$.  }
        \label{pic:sqrtg_plotmcc4c5}
\end{figure}

Now that we have the conjectured analytic forms for $c_2$ and $c_3$, we can subtract off their contribution to the curve and attempt to find the next two coefficients.  Fitting to a curve up to $g^{-5}$, we find that $c_4=-0.008196\pm 3\times 10^{-6}$ and $c_5=-0.00671\pm 3\times 10^{-5}$.  The resulting curve, along with the curve  $c_4g^{-3/2}+c_5g^{-2}$ are shown in figure \ref{pic:sqrtg_plotmcc4c5}\,.  It would be interesting to find the corresponding analytic results.

\section*{Acknowledgements}

We thank J. Maldacena for an explanation of his unpublished results. 
This research is supported in part by
the Swedish Research Council under grant \#2020-03339 and by the National Science Foundation under Grant No.~NSF PHY-1748958. Computations were done on a  cluster provided by the National Academic Infrastructure for Supercomputing in Sweden (NAISS) at UPPMAX, partially funded by the Swedish Research Council under grant \#2022-06725.
J.A.M. thanks the Center for Theoretical Physics at MIT and the KITP for 
hospitality during the course of this work.


\bibliographystyle{JHEP}
\bibliography{hagedorn}  
 
\end{document}